# Nickel Titanium Alloy failure analysis under thermal cycling and mechanical Loading: A Preliminary Study

Mahdi Mohajeri
Department of Materials Science and Engineering
Texas A&M University
College Station, Texas, 77840
USA

Behrouz Haghgouyan
Department of Materials Science and Engineering
Texas A&M University
College Station, Texas, 77840
USA

Homero Castaneda-Lopez
Department of Materials Science and Engineering
Texas A&M University
College Station, Texas, 77840
USA

Dimitris C. Lagoudas
Department of Materials Science and Engineering
Texas A&M University
College Station, Texas, 77840
USA

## ABSTRACT

The electrochemical frequency modulation (EFM) technique can consider as a new tool for electrochemical corrosion monitoring. The calculation of corrosion rate with a non-destructive and rapid technique is a necessity to study corrosion behavior of metals under loading and thermal cycling. NiTi shape memory alloy (SMA) is characterized by differential scanning calorimetry (DSC) and uniaxial tensile testing. The corrosion behavior and reliability of technique have been examined for NiTi sample in artificial physiological solution. The results show the sensitivity of EFM technique to temperature and base frequencies.

## INTRODUCTION

Shape memory alloys (SMAs) are unique class of materials that undergo martensitic phase transformation. This is a solid to solid, diffusenless phase transformation between austenite and martensite, and can be triggered by a change in load and/or temperature. Because of their unique properties, SMAs have wide range of applications [1]. High energy density, makes them a promising material for actuation applications [2]. NiTi, a nearly equiatmoic alloy of nickel and titanium, is the mostly known SMA. Introducing constitutive models [3, 4], failure of SMAs has been investigated numerically [5] and experimentally [6] . Experimental work on characterizing the corrosion behavior of NiTi SMAs is very limited [7, 8].

The corrosion rates in metal samples are determined by several conventional techniques such as: electrochemical impedance spectroscopy (EIS), linear polarization resistance (LPR) and Tafel extrapolation. LPR and Tafel extrapolation methods are used to study corrosion rate of a system where the system needs to be polarizaed over a wide potential range which affects surface of electrode. On the other hand, EIS technique is considered as a non-destructive method with disadvantage of time consuming. Electrochemical frequency modulation (EFM) is a fast and non-destructive corrosion measurement technique which doesn't need to prior knowledge of Tafel constants.
EFM is a good candidate to study corrosion rate of shape memory alloy instantaneously.

## EXPERIMENTAL PROCEDURE

Equiatomic nickel-titanium (NiTi) SMA is acquired from Allegheny Technologies Incorporated (ATI)[1] in the form of 1.87 mm – thick plate. To obtain the transformation temperatures, differential scanning calorimetry (DSC) is carried out using a Perkin–Elmer Pyris[2] 1 machine at a rate of 10 °C/min within -40 to 250 °C temperature range. For tensile characterization, dog-bone specimen with a gauge length of 60 mm and a width of 9 mm is cut from the plate using electrical discharge machining (EDM). Uniaxial tensile testing is carried out at room temperature in a servo-hydraulic MTS test machine with 100 KN load cell. The test is performed in displacement control with a rate of 0.2 mm/min. Full-field in-plane strain is measured using digital image correlation (DIC). The details of the DIC technique are explained elsewhere [9, 10]. Images are recorded by a Point Grey CCD camera fitted with a Tokina AT-X PRO[3] lens and the images are post-processed using Vic-2D[4] software.

A conventional electrochemical cell (Avesta Cell) was used to perform the electrochemical measurements. The experiments were carried out in deaerated artificial physiological solution (0.01 M KCl, 0.05 M CaCl2, 0.15 M NaHCO3, and 0.15 M NaCl) at several of temperatures. The non-destructive EFM technique is used to measure the corrosion rate at each temperature. The potential perturbation signal with amplitude of 10mV vs $E_{corr}$ is used for perturbation frequencies. Two base frequencies were 0.1 Hz and 0.01 Hz with multiplier values of 2 and 5.

## RESULTS

**Material Characterization**

DSC result is illustrated in **Figure 1** according which the transformation temperatures are obtained as $A_s$=88°C, $A_f$=109°C, $M_s$=78°C, $M_f$=51°C where $A_s$, $A_f$, $M_s$, $M_f$ are austenite start, austenite finish, martensite start and martensite finish temperatures, respectively.

---

[1] Allegheny Technologies Incorporated (ATI) Corporate Headquarters 1000 Six PPG Place Pittsburgh, PA 15222-5479 U.S.A. 412-394-2800 ATImetals.com
[2] PerkinElmer Life and Analytical Sciences 710 Bridgeport Avenue Shelton, CT 06484-4794 U.S.A.  203-925-4600 www.perkinelmer.com
[3] Kenko Tokina Co.,Ltd. KT Nakano Building, 5-68-10 Nakano, Nakano-ku, Tokyo 164-8616, Japan  3-6840-3037
[4] Correlated Solutions, Inc. 121 Dutchman Blvd. Irmo, SC 29063 U.S.A. 803-926-7272

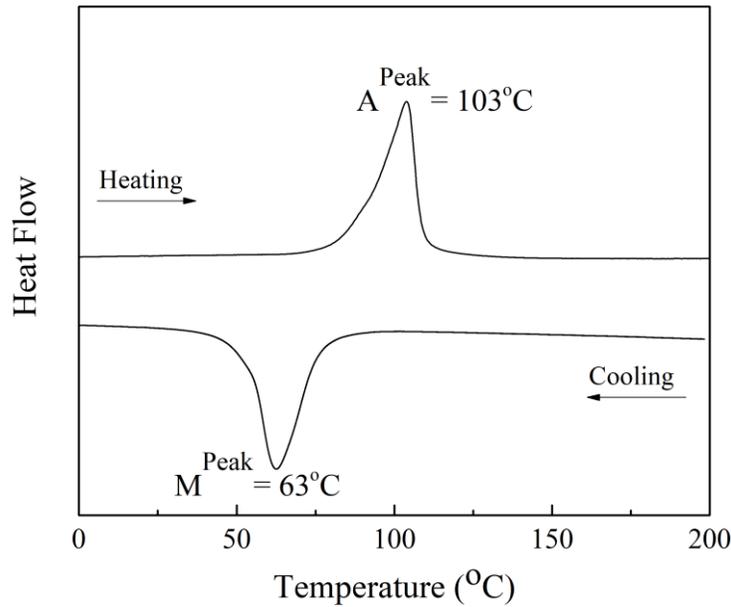

**Figure 1. DSC result of NiTi SMA.**

**Figure 2** shows the stress-strain curve of NiTi at room temperature. According to the DSC results, room temperature is below $M_s$, i.e. the material is in fully martensite state. The deformation mechanism of the material can be explained by dividing the stress-strain curve into four regions. In the first region (deformation up to ~1% strain), the twinned martensite is elastically deformed. Small deviations from linearity in this region is due to the start of detwinning process. In the second region (~1 – 4% strain), a plateau is observed at a stress of ~240 MPa. In this region the deformation is dominated by detwinning of martensite variants. The elastic modulus of martensite is obtained from the slope of unloading line to be ~40 GPa. In the third region (~4 – 10% strain), the martensite that is fully detwinned deforms elastically. Finally, in the last region (~10 – 60% strain), deformation is dominated by slip plasticity and the specimen fails at an ultimate tensile strength of ~900 MPa.

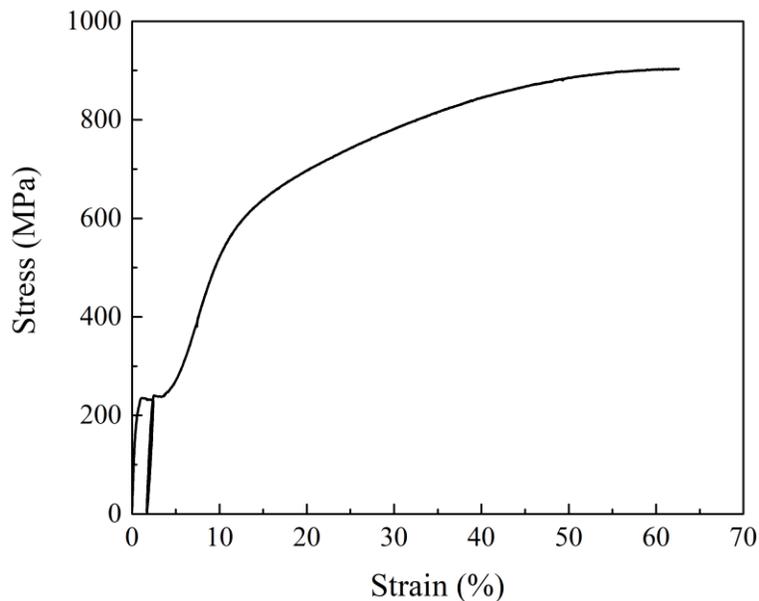

**Figure 2. Stress-strain response of NiTi at room temperature.**

**Figure 3** shows the strain distribution in the loading direction, $\varepsilon_{yy}$, obtained from DIC. The contour plots correspond to applied strain values from 1% to 4% where the martensitic NiTi undergoes elastic deformation followed by detwinning. The strain distribution is nearly uniform in the beginning i.e. in the elastic regime.

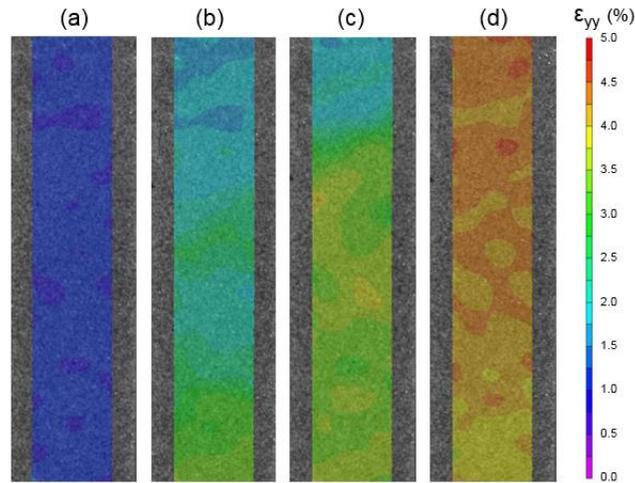

**Figure 3. Strain distribution in loading direction ($\varepsilon_{yy}$) corresponding to (a) 1%, (b) 2%, (c) 3%, and (d) 4% applied strain.**

Strain localization with crossing pattern starts to appear when the material undergoes detwinning. The localization is more pronounced in contour plots of **Figure 4** where the specimen experiences higher strain values. Note that the color bar is reconfigured for better demonstration of strain distribution.

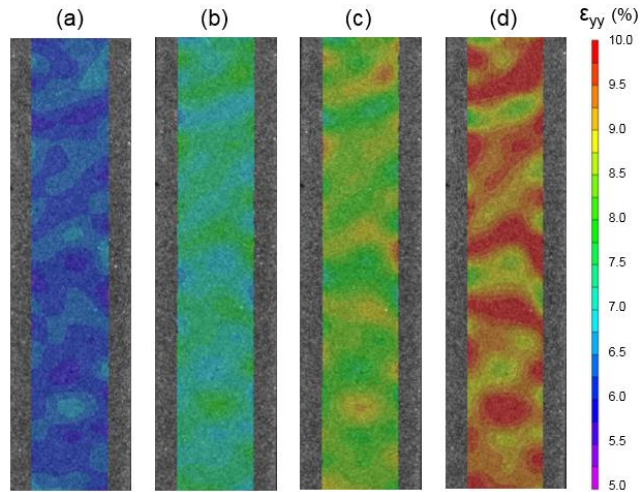

**Figure 4. Strain distribution in loading direction ($\varepsilon_{yy}$) corresponding to (a) 6%, (b) 7%, (c) 8%, and (d) 9% applied strain.**

**Corrosion rate monitoring**
EFM technique is fast and nondestructive method to measure the corrosion rate instantaneously. Tafel parameters are found with a potential perturbation consisting of sine waves of different frequencies with a small polarization. The potential perturbation is defined as follow:

$$\eta = U_0 \sin \omega_1 t + U_0 \sin \omega_2 t \quad \text{Equation(1)}$$

where $\eta$ is the overpotential, $U_0$ the amplitude of the potential perturbation (in this experiment $U_0$=10 mV), and $\omega_1$ and $\omega_2$ are the frequencies[11].
Generated current responses are at more frequencies than applied signals because of the nonlinear nature of corrosion process. Figure 5 shows the applied frequency spectrum for base frequency 0.1 Hz and 0.01 Hz and with amplitude of 10 mV and system responses in time domain and frequency domain at 90°C temperature.

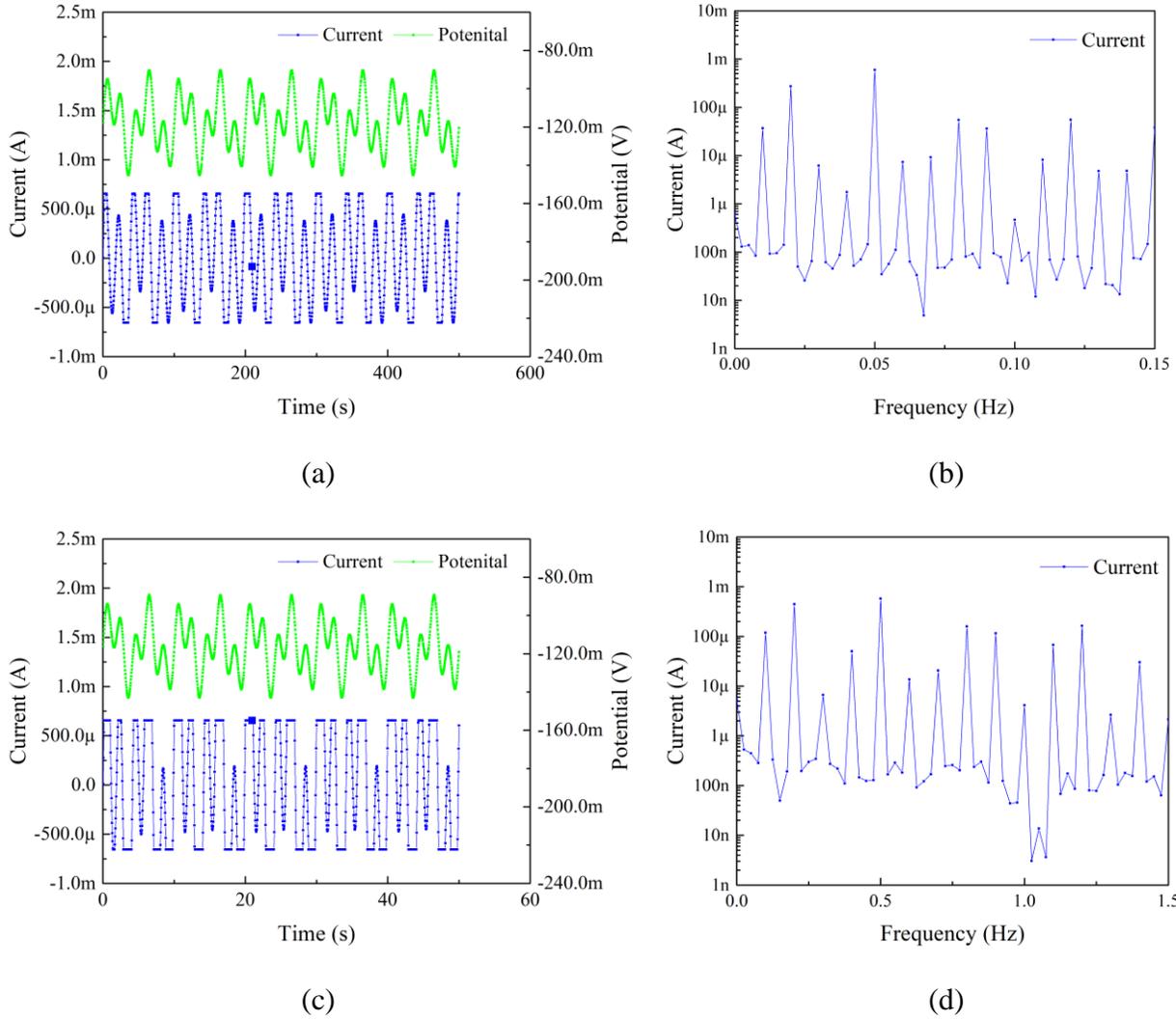

(a)  (b)  (c)  (d)

**Figure 5- (A) Potential perturbation and current responses with (a) 0.01Hz and (c) 0.1 Hz base frequency in time domain, Frequency spectrum of the current response with (b) 0.01Hz and (d) 0.1 Hz base frequency in frequency domain for the NiTi in artificial physiological solution at 90°C.**

The corrosion current density and Tafel parameters can be defined by mathematical analysis of current components in frequency domain with the following relations[11, 12]:

$$i_{corr} = \frac{i_{\omega_1,\omega_2}^2}{2\sqrt{8 i_{\omega_1,\omega_2} i_{2\omega_2 \pm \omega_1} - 3 i_{\omega_2 \pm \omega_1}^2}}$$

Equation(2)

$$\beta_a = \frac{i_{\omega_1,\omega_2} U_0}{i_{\omega_2 \pm \omega_1} + \sqrt{8 i_{\omega_1,\omega_2} i_{2\omega_2 \pm \omega_1} - 3 i_{\omega_2 \pm \omega_1}^2}}$$

Equation(3)

$$\beta_c = \frac{i_{\omega_1,\omega_2} U_0}{-i_{\omega_2 \pm \omega_1} + \sqrt{8 i_{\omega_1,\omega_2} i_{2\omega_2 \pm \omega_1} - 3 i_{\omega_2 \pm \omega_1}^2}}$$

Equation(4)

where $i_{corr}$ is he corrosion current and $\beta_a$ and $\beta_c$ are the anodic and cathodic Tafel parameters, respectively. The current responses at harmonic frequencies are $i_{\omega_1,\omega_2}, i_{2\omega_1}, i_{2\omega_2}$ and at intermodulation frequencies are $i_{\omega_2\pm\omega_1}, i_{2\omega_2\pm\omega_1}$. The following relationship exist between the harmonic and intermodulation components of current response:

$$i_{\omega_2\pm\omega_1} = 2i_{2\omega_1} = 2i_{2\omega_2} \qquad \text{Equation(5)}$$

$$i_{2\omega_2\pm\omega_1} = i_{\omega_2\pm 2\omega_1} = 2i_{2\omega_1} = 2i_{2\omega_2} \qquad \text{Equation(6)}$$

These components provide the causality factors:

$$Causality\_factor(2) = \frac{i_{\omega_2\pm\omega_1}}{i_{2\omega_1}} \qquad \text{Equation(7)}$$

$$Causality\_factor(3) = \frac{i_{2\omega_2\pm\omega_1}}{i_{3\omega_1}} \qquad \text{Equation(8)}$$

The causality factors (Figure 6) can verify the quality of the obtained experimental data with EFM technique.

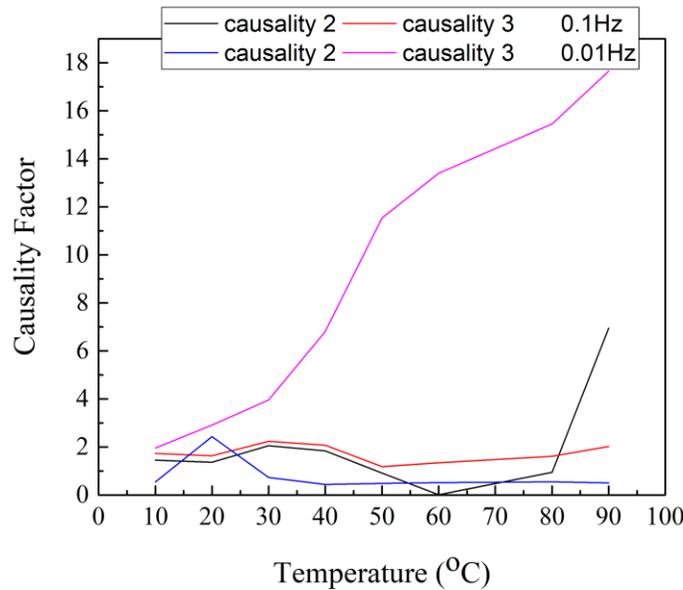

**Figure 6- Influence of the temperature on the causality factors are calculated for base frequencies 0.1 and 0.01 Hz.**

Table 1 shows the different corrosion parameters for NiTi sample during cooling from 0°C to 90°C. The experimental results for corrosion rates show the constant corrosion rate for NiTi when the 0.1Hz base frequency has been used in EFM techniques.

**Table 1- Electrochemical parameters obtained by EFM technique for NiTi at several temperatures in artificial physiological solution.**

| Temperature (°C) | EFM technique in 0.1 Hz base frequency | | | EFM technique in 0.01 Hz base frequency | | |
|---|---|---|---|---|---|---|
| | $i_{corr}$ (μA) | $\beta_a$ (mV/dec) | $\beta_c$ (mV/dec) | $i_{corr}$(μA) | $\beta_a$(mV/dec) | $-\beta_c$(mV/dec) |
| 90 | 173.5 | 21.72 | 22.52 | 235.9 | 34.79 | 36.17 |
| 80 | 175.6 | 22.16 | 23 | 286.3 | 45.23 | 46.52 |
| 70 | 179.3 | 22.95 | 23.78 | 387.8 | 65.87 | 67.17 |
| 60 | 182.5 | 23.68 | 24.45 | 576.7 | 102.7 | 105.8 |
| 50 | 188.4 | 25.05 | 25.66 | 2669 | 517.1 | 552.6 |
| 40 | 198.6 | 27.25 | 27.77 | 2278 | 486.6 | 513.2 |
| 30 | 209.5 | 29.43 | 30.11 | 1672 | 338 | 351 |
| 20 | 241.4 | 36.02 | 36.68 | 1321 | 302.4 | 313.2 |
| 10 | 173.5 | 21.72 | 22.52 | 235.9 | 34.79 | 36.17 |

With 0.01 Hz base frequency, the measured corrosion rates increased with decreasing temperatures from 90°C to 50°C and decreased with decreasing temperature from 50°C to 10°C. This variation in results shows the necessity of data validation for system. Causality factor values in Figure 6 indicates that the measured data with 0.1Hz base frequency have good quality below the 80 °C and data measured with 0.01 Hz base frequency from 10°C to 30°C have acceptable derivation from ideal causality factors.

## CONCLUSIONS

The corrosion rate of NiTi SMA is studied under thermal cycling in physiological solution with non-destructive EFM method. The results show the variation in obtained data quality from good to poor. The data with good quality can be used to evaluate Tafel parameters and corrosion rate of NiTi at each temperature. Similar to thermal cycling experiment, the corrosion rate measurement during mechanical loading (tensile test) needs a non-destructive reliable technique (e.g. EFM) can be used as a technique without prior knowledge of Tafel constants. The obtained data with this technique can be verified with causality factors which is considered as an internal check for validity.